\newif\ifpdf
   \ifx\pdfoutput\undefined
   \pdffalse  % not running pdflatex
\else
   \pdfoutput=1  % running pdflatex
   \pdftrue
\fi

\documentclass[11pt,a4paper]{article}
\ifpdf
\usepackage[pdftex]{graphicx}
\else
\usepackage[dvips]{graphics}
\fi
\usepackage{amsmath,amssymb}

\par
\begin{document}

\begin{title}{Electronic Energy Spectra and Wave Functions\\ on the Square
Fibonacci Tiling}\end{title}
\author{SHAHAR EVEN-DAR MANDEL and RON LIFSHITZ\\
School of Physics and Astronomy, Raymond and Beverly Sackler Faculty\\
of Exact Sciences, Tel Aviv University, Tel Aviv 69978,
Israel}
\date{May 15, 2005}

\maketitle

\begin{abstract}
  We study the electronic energy spectra and wave functions on the
  square Fibonacci tiling, using an off-diagonal tight-binding model,
  in order to determine the exact nature of the transitions between
  different spectral behaviors, as well as the scaling of the total
  bandwidth as it becomes finite. The macroscopic degeneracy of certain
  energy values in the spectrum is invoked as a possible mechanism
  for the emergence of extended electronic Bloch wave functions as the
  dimension changes from one to two.
\end{abstract}

\section{Background and Motivation}

We continue our initial studies~\cite{ilan} of the off-diagonal
tight-binding hamiltonian on the square Fibonacci
tiling~\cite{squarefib}, in order to gain a better quantitative
understanding of the nature of the transitions between different
spectral behaviors in this 2-dimensional ($2d$) quasicrystal. We
also consider more carefully the transition of the spectrum from
singular-continuous to absolutely continuous, in going from one to
two dimensions, and the implications of this transition on the
possible emergence of extended Bloch wave functions.

The square Fibonacci tiling is constructed by taking two identical
grids---each consisting of an infinite set of lines whose
inter-line spacings follow the well-known Fibonacci sequence of
short ($S$) and long ($L$) distances---and superimposing them at a
$90^{\circ}$ angle. This construction can be generalized, of
course, to any quasiperiodic sequence as well as to higher
dimensions. On this tiling we define an off-diagonal tight-binding
hamiltonian with zero on-site energy, and hopping amplitudes: $t$
for vertices connected by short ($S$) edges; $1$ for vertices
connected by long ($L$) edges; and zero for vertices that are not
connected by edges.

The 2-dimensional Schr\"odinger equation for this model is given by
\begin{eqnarray}
  \label{eq:twoDeq}\nonumber
  &t_{n+1} \Psi(n+1,m) + t_n \Psi(n-1,m)\hskip3cm\\
  &+ t_{m+1} \Psi(n,m+1) + t_m \Psi(n,m-1)
  = E\Psi(n,m),
\end{eqnarray}
where $\Psi(n,m)$ is the value of a $2d$ eigenfunction on a
vertex labeled by the two integers $n$ and $m$, and $E$ is the
corresponding eigenvalue. The hopping amplitudes $t_j$ are equal to 1
or $t$ according to the Fibonacci sequence as described above, and we do
not lose much by limiting ourselves to the case $t>1$, as the
important features of the model depend on the ratio of the two
hopping amplitudes.

With no additional assumptions other than the absence of diagonal
hopping this 2-dimensional eigenvalue problem, as well as its higher
dimensional versions, are all separable. This allows us to use the
known solutions for the $1d$ problem~\cite{1dmodels,reviews} in order
to construct the solutions in two and higher dimensions (as was done
for similar models in the past~\cite{2dmodels,ashraff}).
Two-dimensional eigenfunctions can therefore be expressed as products
\begin{equation}
  \label{eq:product}
  \Psi_{ij}(n,m)=\psi_i(n)\psi_j(m),
\end{equation}
with eigenvalues
\begin{equation}
  \label{eq:sum}
  E_{ij}=E_i + E_j,
\end{equation}
where $\psi_i(n)$ and $\psi_j(n)$ are two of the eigenfunctions of the
corresponding $1d$ eigenvalue equation on the Fibonacci chain, with
eigenvalues $E_i$ and $E_j$. Consequently, the $2d$ density of states
(DOS) is given by a convolution of two $1d$ densities of states.

\section{Study of the $2d$ spectrum}

Although the spectrum of the $1d$ model is always a Cantor-like
set with zero bandwidth and an infinite number of bands, the $2d$
spectrum exhibits different behavior for different ranges of the
parameter $t$. For $t$ close to 1 the gaps in the $1d$ spectrum
are small and disappear as a result of the addition of
spectra~(\ref{eq:sum}), producing an absolutely continuous $2d$
spectrum. For $t\gg1$ the $2d$ spectrum retains the Cantor-like
structure of the $1d$ spectrum. For intermediate values of $t$ the
number of bands tends to infinity but the total bandwidth remains
finite and it is unclear whether the spectrum has an absolutely
continuous part.

In order to study the exact structure of the $2d$ spectrum---the total
bandwidth and number of bands as a function of $t$---one has to
perform a convolution of two $1d$ spectra. The numerical study of the
$1d$ spectrum is difficult for low values of $t$ ($t\approx 1$) for
which the gaps are very small, and for $t\gg 1$ for which the bands
are very small. Further numerical difficulties are encountered in the
convolution process itself. The desired $2d$ spectrum, consisting of
the set of points for which the $2d$ DOS is not zero, is given by
\begin{equation}
\label{eq:2dspectrum}
\bigcup_{i,j=1}^{F_N}\left[a_i^{(N)}+a_j^{(N)},b_i^{(N)}+b_j^{(N)}\right],
\end{equation}
where $\left[a_i^{(N)},b_i^{(N)}\right]$ is the $i^{th}$ interval in
the relevant $1d$ spectrum of the $N^{th}$ approximant, and $F_N$ is
the $N^{th}$ Fibonacci number. Because $F_N\sim \tau^N$ the number of
intervals in the union~(\ref{eq:2dspectrum}) behaves like $\tau^{2N}$,
and although many of the intervals may overlap, any attempt to
directly study the band structure and the DOS requires a construction
of the whole set. The amount of memory required to store such sets
makes exact calculation implausible for high values of $N$ regardless
of the value of $t$. Luckily, the total bandwidth of the $2d$ spectrum
can be calculated without full knowledge of the DOS. This is done by
examination of a set of uniformly distributed energies in the relevant
range and testing for each energy whether it lies within one of the
above intervals. For a sufficiently large set of energies this
procedure yields the measure of the spectrum after proper
normalization.

%##################### Figure 1####################################
\begin{figure}[tb]
\hspace*{0pt}
\begin{center}
  \scalebox{0.9}{\rotatebox{00}{\includegraphics*{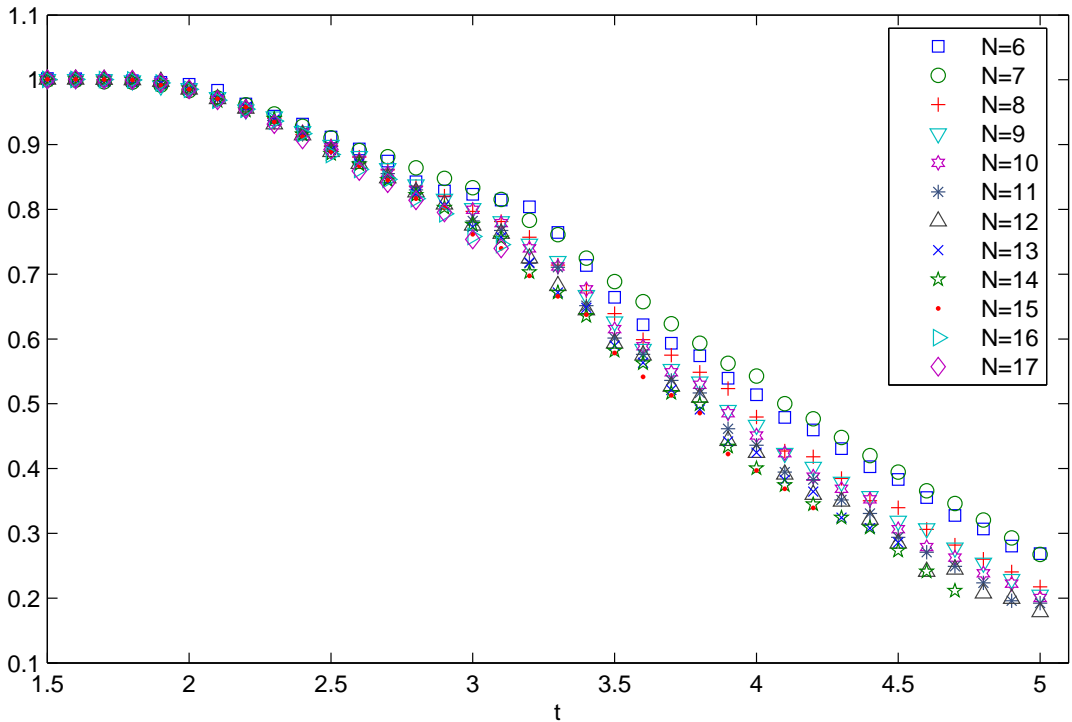}}}\\
  \caption{Total bandwidth of the $2d$ spectrum
    divided by $4r_N(t)$. For values of $t$ up to $\sim2$ the spectrum
    consists of a single band regardless of $N$. At higher values the
    bandwidth clearly depends on $N$.
\label{fig:bandwidth}}
\end{center}
\end{figure}
%##################### Figure 1####################################

%##################### Figure 2####################################
\begin{figure}[tb]
\hspace*{0pt}
\begin{center}

\scalebox{0.8}{\rotatebox{00}{\includegraphics*{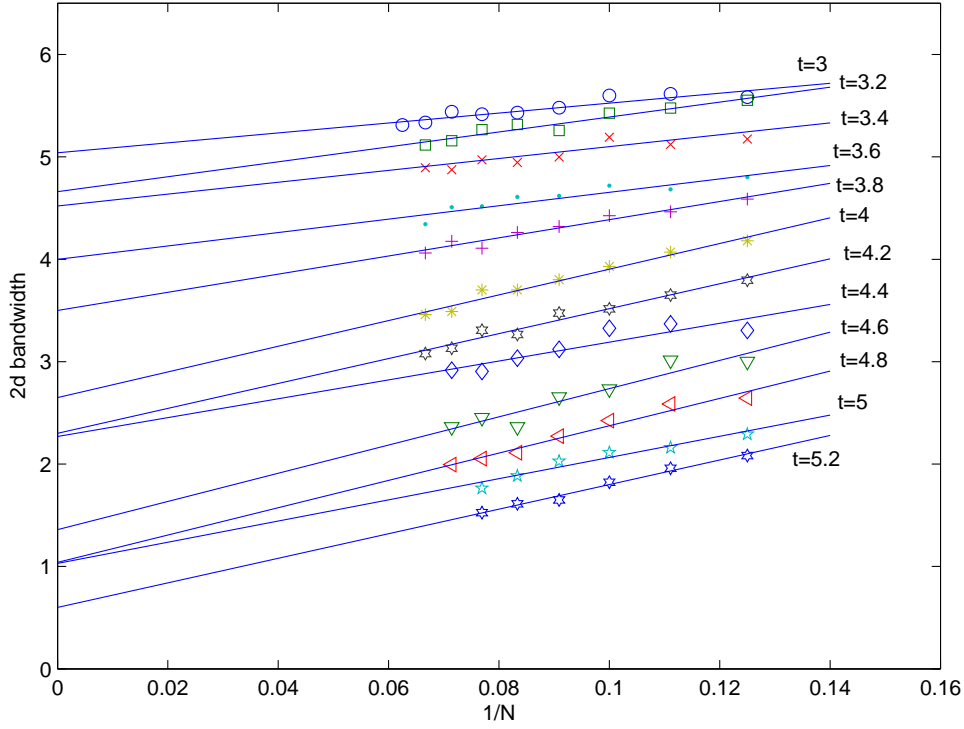}}}\\
\caption{Total bandwidth as function of $1/N$ for different values of
  $t$. Linear extrapolation to $1/N\to0$ gives a prediction for the
  total bandwidth at the infinite $N$ limit.
\label{fig:finitesize}}

\end{center}
\end{figure}
%##################### Figure 2####################################

\begin{figure}[tb]
\hspace*{0pt}
\begin{center}

\scalebox{0.8}{\rotatebox{00}{\includegraphics*{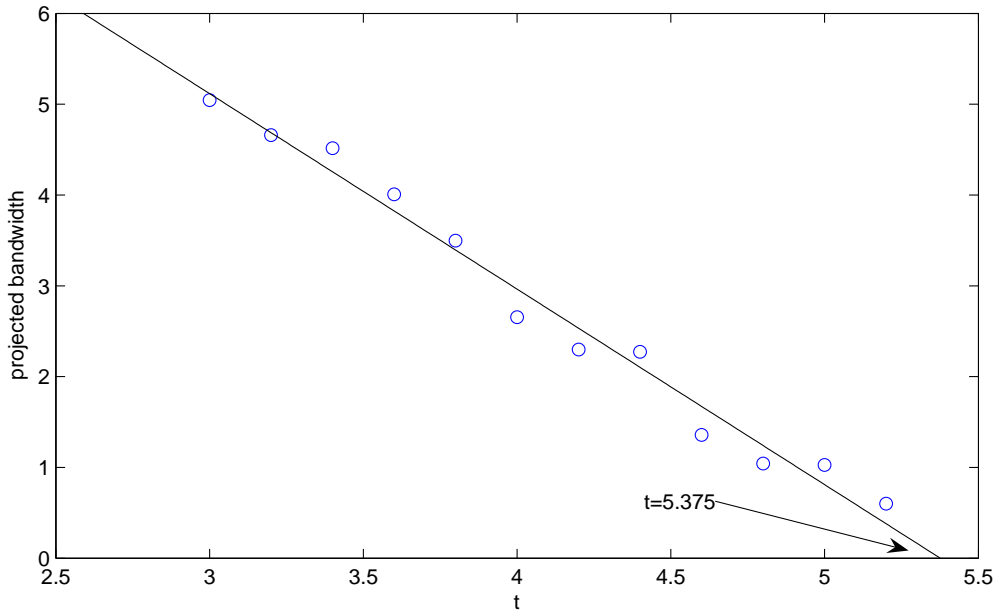}}}\\
\caption{Extrapolation of the predicted infinite
  limit bandwidth to find the critical value $t_c$, above which the
  total bandwidth tends to zero.
\label{fig:tc}}

\end{center}
\end{figure}

For given values of $N$ and $t$ the $1d$ spectrum is a symmetric set
of $F_N$ intervals within $\left[-r_N(t),r_N(t)\right]$, where
$r_N(t)$ is the maximal eigenenergy of the $N^{th}$ $1d$ approximant.
The $2d$ spectrum is therefore a subset of the interval
$\left[-2r_N(t),2r_N(t)\right]$, which gives an upper bound of
$4r_N(t)$ on its total bandwidth. Fig.~\ref{fig:bandwidth} shows the
total bandwidth divided by $4r_N(t)$ as a function of $t$ for a
sequence of approximants $N=6\ldots 17$.  For values of $t$ up to
$\sim2$ the spectrum consists of a single band. For $t$'s above 2 the
normalized bandwidth becomes dependent on $N$.  Using finite-size
scaling (as suggested by Ashraff {\it et al.}~\cite{ashraff}) we can
obtain an estimate for the critical value $t_c$ at the transition
between the intermediate and large $t$ regimes, where the total
bandwidth vanishes. Fig.~\ref{fig:finitesize} shows the total
bandwidth as a function of $1/N$ for different values of $t>2$.
Though some oscillatory behavior is observed, the behavior seems
quite linear and extrapolation to $1/N\to0$ yields a prediction for
the total bandwidth in the infinite $N$ limit.  By plotting these
estimates as a function of $t$ we obtain Fig.~\ref{fig:tc} which
again exhibits an overall linear behavior. Extrapolation yields an
estimate for a critical value of $t_c\simeq 5.375$ for the transition
between the intermediate and large $t$ regimes.

%##################### Figure 3####################################
%##################### Figure 2####################################

It is worth noting that as $t$ is increased and the gaps grow, the
degeneracy of energy levels decreases. At the asymptotic limit
each interval in the $2d$ spectrum should be exactly doubly
degenerate (neglecting a single interval around $E=0$). When this
limit is obtained the $2d$ bandwidth should be approximately equal
to $F_NB_{1d}(t,N)$, where $B_{1d}(t,N)$ is the bandwidth of the
corresponding $1d$ spectrum.  The numerical data shows that this
limit is to be obtained well beyond $t_c$. It is hence evident
that the large $t$ regime of the $2d$ spectrum is further divided
into two regimes, one (at extremely large $t$) in which the energy
levels are no-more than doubly degenerate, and another (at
moderately large $t$) in which energy levels may have a higher
degeneracy.

\section{Emergence of extended wave functions}

Since the eigenfunctions of the $1d$ hamiltonian are critically
localized it is clear that the $2d$ eigenfunctions~(\ref{eq:product})
should also be critically localized (Fig. 4(a-c)). However, the
existence of a regime for $t$ close to 1 in which the spectrum, or at
least a part of it, is absolutely continuous hints at the existence
of extended wave functions. We proposed in the past~\cite{ilan} that
the degeneracy of the $2d$ spectrum, at values of $t$ close to 1, may
provide an explanation for the emergence of extended wave functions
as the dimensionality increases.

%##################### Figure 4 ####################################
\begin{figure}[thp]
\hspace*{0pt}
\begin{center}
\begin{tabular}{c c c}
\scalebox{0.73}{{\includegraphics{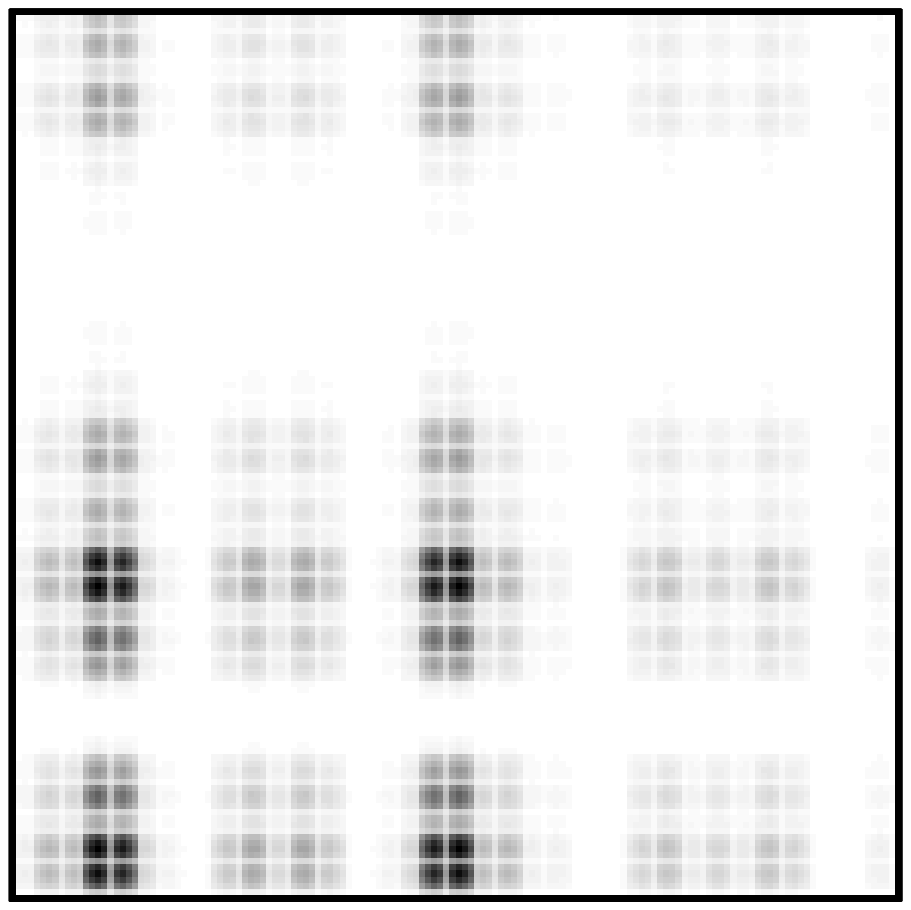}}}  & \hphantom{----.------} &
\scalebox{0.73}{{\includegraphics{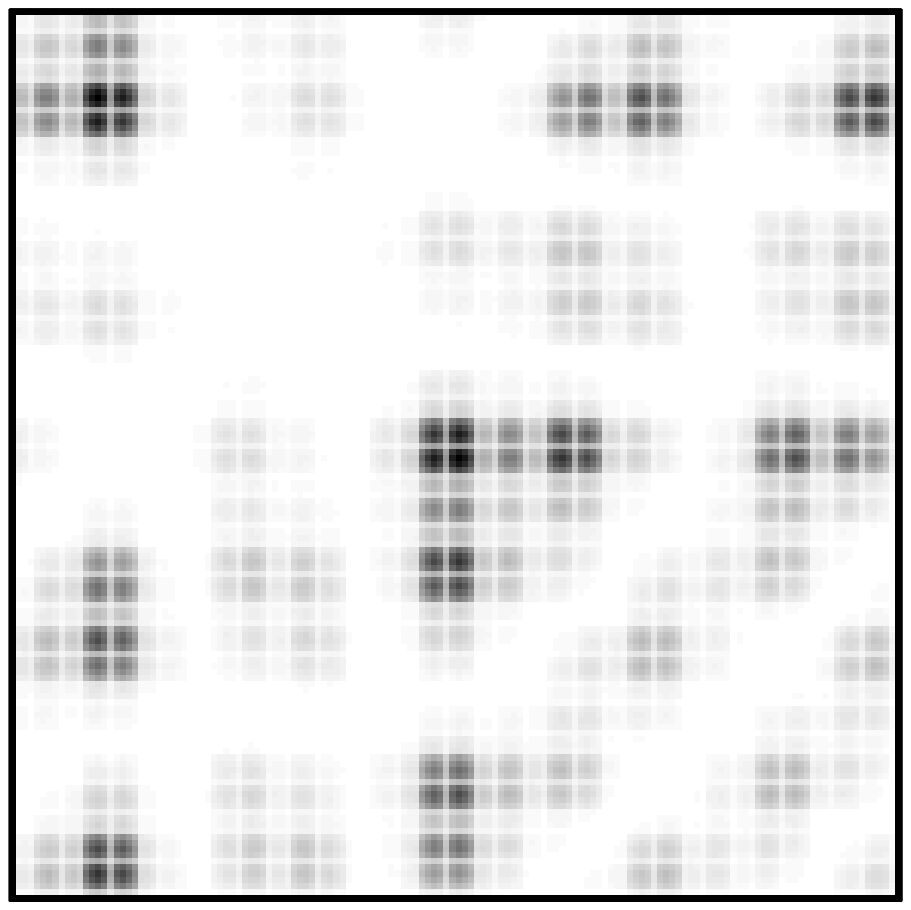}}} \\
(a) & & (b) \\
\scalebox{0.73}{{\includegraphics{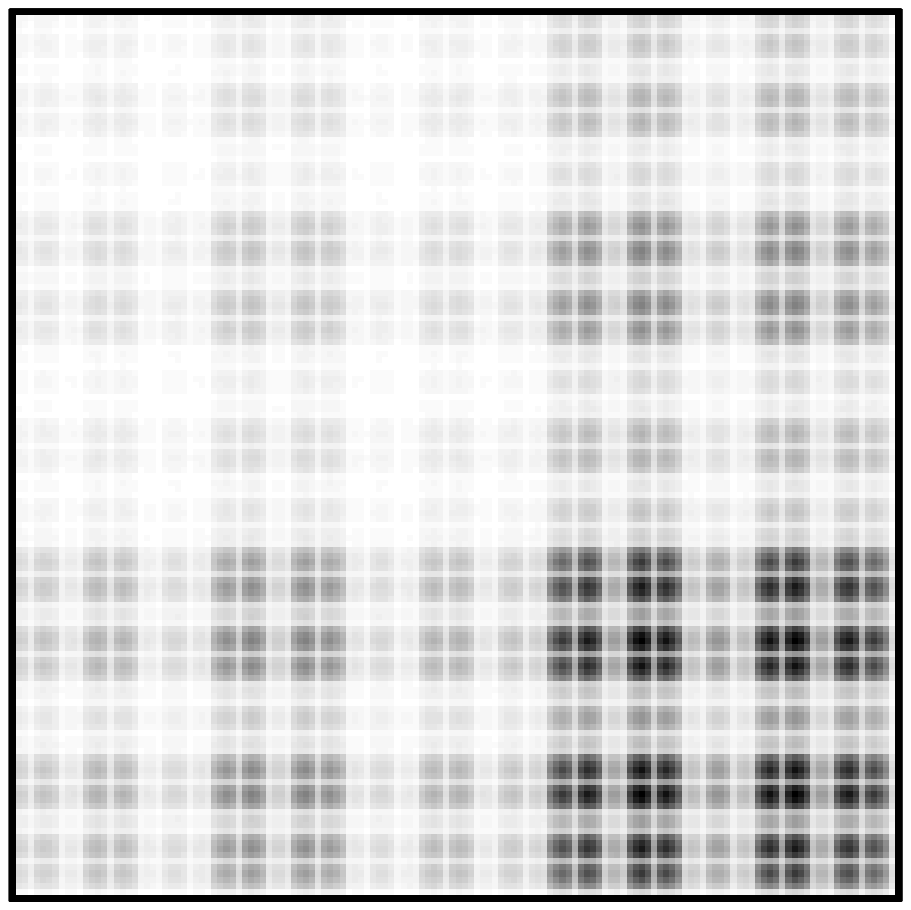}}} & &
\scalebox{0.73}{{\includegraphics{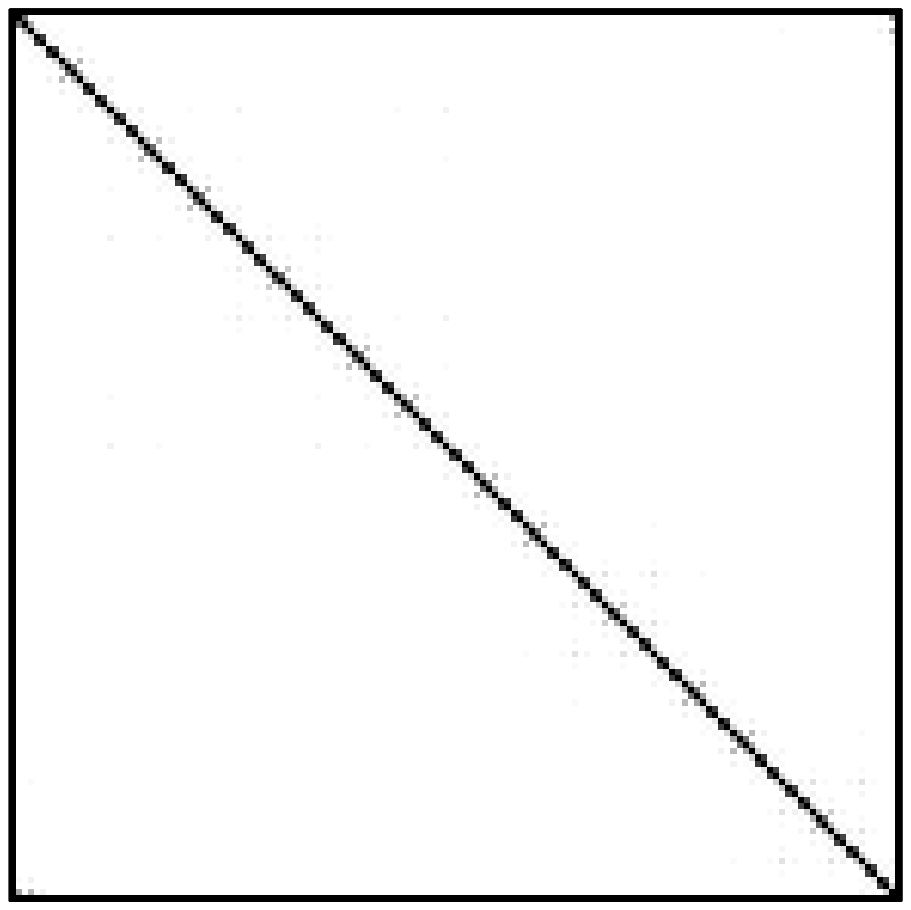}}} \\
(c) & & (d)\\
\end{tabular}

\caption{Wave functions on a single $144\times144$ $2d$ approximant.
(a) A critical wave function $\Psi_{n,m}$; (b) A symmetrized critical
wave function $\frac{1}{\sqrt{2}}\left(\Psi_{n,m}+\Psi_{m,n}\right)$;
(c) A zero energy wave function $\Psi_{-n,n}$; and (d) A linear
combination of zero energy wave functions producing an extended Bloch
wave function along the diagonal, clearly showing the underlying
Fibonacci modulation.
\label{fig:wavefns}}

\end{center}
\end{figure}
%##################### Figfure 4 ####################################

To illustrate the possible role of degeneracy in the emergence of
extended wave functions we analyze the $E=0$ eigenfunction. These
functions are macroscopically-degenerate for any value of $t$. A set
of $1d$ eigenfunctions can be found by directly diagonalizing the
$1d$ hamiltonian for a single approximant with periodic boundary
conditions. The $F_N$ eigenfunctions obtained in this way generate
$F_N^2$ eigenfunctions for the $2d$ model, $F_N$ of which have $E=0$.
These eigenfunctions, given by
$\Psi_i(n,m)=(-1)^m\psi_i(n)\psi_i(m)$, are each critically
localized, but we can demand that a linear combination
$\sum_{i=1}^Na_i\Psi_i(n,m)$ of them have certain values on a given
subset of sites by solving the equations for the coefficients $a_i$.
Setting the values of a linear combination to be identical on half
the points along the diagonal results in eigenfunctions of the form
shown in Fig.~\ref{fig:wavefns}(d). This wave-function is an extended
Bloch function along the diagonal---clearly showing the underlying
Fibonacci modulation---but is strongly localized in the perpendicular
direction.

Whether the degeneracy of other energy values is sufficiently high
to allow for a similar process to be carried out is not clear, but
the chances for success should strongly improve as the tunnelling
amplitude $t$ approaches 1. This also corresponds to the regime in
which the spectrum is (at least partly) absolutely continuous. For
large $t$, where the spectrum is singular continuous, the
degeneracy of levels other than $E=0$ is very small and no
extended wave functions are expected to be found.
Note that in 3 dimensions all the energies of the original $1d$
spectrum are macroscopically degenerate. This means that extended
wave-functions, similar to that of Fig.~\ref{fig:wavefns}(d), can be
constructed for a wider range of energies. Thus, as the
dimensionality further increases so does the degeneracy of energy
levels, and with it the ability to construct Bloch wave functions.

\noindent {\bf Acknowledgments}

This research is supported by the Israel Science Foundation
through Grant No.~278/00.

\end{document}